\documentclass[10pt,twocolumn,showpacs,superscriptaddress,notitlepage,pre,aps,amsmath,amssymb,preprintnumbers]{revtex4}
\usepackage{epsfig}
\usepackage{graphicx}
\usepackage{dcolumn}

 \let\r=\rho \let\th=\theta \let\io=\infty

\def\to{\rightarrow}

\newcommand{\beq}{\begin{equation}} \newcommand{\eeq}{\end{equation}}

\begin{document}

\title{Application of Edwards' statistical mechanics to high dimensional
jammed sphere packings}
\author{Yuliang Jin}
\affiliation{Levich Institute and Physics Department, City College
of New York, New York, New York 10031, USA}
\author{Patrick~Charbonneau}
\homepage{http://www.chem.duke.edu/labs/charbonneau/}
\affiliation{Department of Chemistry and Department of Physics, Duke University, Durham,
North Carolina 27708, USA}
\author{Sam Meyer}
\affiliation{Levich Institute and Physics Department, City College
of New York, New York, New York 10031, USA}
\affiliation{Laboratoire de physique de l'ENS Lyon, CNRS UMR 5672,
Universit\'e de Lyon, Lyon, France}
\author{Chaoming Song}
\affiliation{Levich Institute and Physics Department, City College
of New York, New York, New York 10031, USA}
\affiliation{Center for Complex Network Research, Departments of
Physics, Biology, and Computer Science, Northeastern University,
Boston, Massachusetts, 02115, USA}
\author{Francesco Zamponi}
\affiliation{LPTENS, CNRS UMR 8549, associ\'ee \`a l'UPMC Paris
06, 24 Rue Lhomond, 75005 Paris, France}
\date{\today}

\begin{abstract}
The isostatic jamming limit of frictionless spherical particles
from Edwards' statistical mechanics [Song \emph{et al.}, Nature (London) {\bf 453}, 629 (2008)] is
generalized to arbitrary dimension $d$ using a liquid-state description.
The asymptotic high-dimensional behavior of the self-consistent relation is obtained by
saddle-point evaluation and checked numerically. The resulting
random close packing density scaling $\phi\sim
d\,2^{-d}$ is consistent with that of other
approaches, such as replica theory and density functional theory. The validity of various
structural approximations is assessed by comparing with three- to six-dimensional isostatic packings obtained from simulations.
These numerical results support a growing accuracy of the theoretical approach with dimension. The approach could thus serve as a starting point to obtain a geometrical understanding of the higher-order correlations present in jammed packings.
\end{abstract}

\maketitle

\section{Introduction}

The sphere packing problem in large spatial dimension $d$ is related to several
important mathematical problems
in the context of signal digitalization and of error correcting codes, in particular.
It has been investigated in detail by the information theory
community \cite{conway:1999,Rogers},
but in spite of such strong interest,
the known rigorous bounds on packing fractions $\phi$ are not very restrictive.
For the lower bound, the classical Minkowsky result
$\phi \sim 2^{-d}$~\cite[Chap. 1,
Sec. 1.5]{conway:1999}
can be improved for lattice packings $\phi \geq 2 d \, 2^{-d}$~\cite{Ba92}, and Ref.~\cite{KLV04} discusses
a procedure to actually construct packings that achieve this bound.
For the upper bound, Kabatiansky and Levensthein have obtained an
asymptotic scaling $\phi \sim 2^{-0.5990\ldots \, d}$~\cite{KL78}. Though the $\phi$ values of laminated lattices up to $d=50$ seem to suggest that there exist lattices where $2^d \phi$ grows exponentially with $d$~\cite[Chap. 6]{conway:1999},
it is quite possible for this observation to result from pre-asymptotic effects.
The gap between the known upper and lower bounds thus grow exponentially with $d$.
This broad uncertainty leaves open the possibility that the densest packings for $d\to \io$ may be amorphous.
It has indeed been proposed in Ref.~\cite{torquato:2006} that
it could be possible to construct amorphous packings that have a density not only exponentially higher
than the Minkowsky lower bound, but actually very close to the Kabatiansky-Levensthein upper bound. A better understanding of high-dimensional amorphous and lattice packings would help clarify this intriguing issue.

Dense amorphous packings of hard spheres are produced according to a specific
dynamical protocol. Typically, one starts from an initial random configuration of spheres
obtained, e.g., by throwing them into a container, then shaking, tapping, or agitating them until a jammed structure is obtained~\cite{SK69,Be72,PV86,Torquato,SGS05,DB06,DMB05,AD06,PNC07,JSSSSA08,MSLB07}.
In numerical simulations, amorphous packings are produced by inflating hard particles while
avoiding superposition via molecular dynamics~\cite{LS90,DTS05,skoge:2006}, by compressing deformable particles~\cite{Makse2000,zhang:2005},  or by minimizing the interaction energy of
soft particles~\cite{CJ93,OLLN02,OSLN03,SLN06,SHESS07}.
It is an observational fact that
these procedures, when crystallization is avoided,
lead to a final packing fraction
close to $0.64$ in $d=3$.
This ``random close packing density'', which is approximately $10\%$ smaller than the density of the best lattice packing in the same
dimension, is conjectured to be the densest
possible packing that does not display local crystalline order. An important remark for the
following discussion is that the random close packings are found to be ``isostatic'', that is, their average
coordination number $z$ is at the limit of mechanical stability $z=2 d$~\cite{OLLN02,OSLN03,DTS05,skoge:2006}.
A completely
satisfactory characterization of the amorphous states of a system of
identical hard spheres is, however, not yet available. The definition of amorphous close
packed states is still matter of debate~\cite{TTD00,OLLN02,KL07,jin:2010,radin:2008}, in part because the metastability of the jammed amorphous state with respect to the crystal order leads to a thermodynamic ambiguity~\cite{PZ10}.

Classical statistical mechanics provide useful insights into the problem of sphere packing in large $d$,
for both lattice
(see Ref.~\cite{Pa07} for an attempt in this direction) and amorphous
packings~\cite{PZ10}.
In the limit of large spatial dimension
mean-field theory becomes exact because each degree of freedom interacts with a large
number of neighbors~\cite{ParisiBook}.
Additionally, because surfaces and volumes scale the same way for large $d$ and geometrical frustration between the liquid and the crystal persists~\cite{vanmeel:2009a,vanmeel:2009},
nucleation is strongly suppressed. Metastability effects become less important, which reduces the definitional ambiguity.
It is thus likely for statistical mechanics to provide precise
information on the behavior of amorphous packings in this limit.

Edwards proposed a volume ensemble statistical approach to study the ``out-of-equilibrium'' nature of jammed states~\cite{edwards:1989,Blumenfeld2003,Srebro2003,Blumenfeld2006,Aste2008,Frenkel2008}, and a simple mean-field
theory based on this approach was recently developed~\cite{song:2008,vanmeel:2009,danisch:2010,meyer:2010}.
At the theory's core is the distribution of
Voronoi volumes associated with particles. Information about the packings can be extracted from the probability distribution of this ``volume function''.
The self-consistent integral equation for the
free volume that results can then be solved
analytically or numerically to derive a relation between the
packing fraction and the average coordination number.
This approach has been used in low dimensions to predict
the density and other properties of jammed packings.

In this paper,
we provide an alternative derivation of the probability distribution of the volume function that is based on a large $d$ approach.
From the theory, we derive a general relation between the density of jammed packings
and their average coordination number $\phi \sim z/2^d$,
that holds at exponential order in $d$ and is consistent
with the one found in Ref.~\cite{torquato:2006} using a different approach.
For isostatic random close packings we then obtain
$\phi \sim (4/3)\,d\,2^{-d}$, which is denser than the Minkowsky lower bound. Comparing the theoretical predictions with computer-generated amorphous packings shows that, though
generally poor, the agreement nonetheless increases with dimension.

The paper is organized as follows. In Sec.~\ref{sec:details} we detail the notation and computer simulation approach. In Sec.~\ref{sec:method} we introduce theoretical method and provide a liquid-state derivation of the self-consistent closure. The saddle-point approximate solution for the high-dimensional limit is given in Sec.~\ref{sect:analphi}. Readers uninterested by the technical details should skip to Subsec.~\ref{subsect:solution} and~\ref{subsect:finiteD}, where the result and its physical interpretation are given. In Sec.~\ref{sec:lowd} we analyze the low-dimensional corrections to the result, and discuss the implications for  $d=3$. We conclude by comparing our results with other theoretical approaches (Sec.~\ref{sec:comp}) and discuss possible improvements to the theory (Sec.~\ref{sec:conclusion}).

\section{Notation and Simulation details}
\label{sec:details}

Following standard mathematical notation, we define the $(d$$-$$1)$-dimensional
volume, i.e., the surface, of the $(d$$-$$1)$-dimensional unit sphere
at the boundary of the $d$-dimensional unit ball
\begin{equation}
S_{d-1}=\frac{2\pi^{d/2}}{\Gamma(d/2)}
\end{equation}
which is related to the volume of the unit ball $V_d=S_{d-1}/d$. The discussion that follows will consider
hard spherical particles of radius $a = 1/2$ and volume $V_g = V_d/2^d$, using the particle diameter as unit of length.
We denote by $\rho = N/V$ the number density of particles and by $\phi= \r V_g$ the packing fraction.

Isostatic random packings in $d=3$$-$$6$ are generated using the simulation code of Skoge~\emph{et al.}~\cite{skoge:2006}
for $N=500$ particles.
The event-driven molecular dynamics simulations use a modified Lubachevsky-Stillinger algorithm to generate jammed
hard-sphere packings. The system dynamically evolves according to Newtonian mechanics until a diverging pressure is obtained.
For sufficiently large compression rates $\gamma$ the compressed fluid falls out of equilibrium, which results in a jammed configuration. Structural analysis reveals that these jammed configurations are isostatic and disordered without any sign of crystallization~\cite{skoge:2006}. We find that compressing the system with $\gamma = 10^{-3}$ in reduced units, until the reduced pressure
$Z\equiv\beta p/\rho = 10^{12}$, with the thermal energy $1/\beta$ set to unity, is sufficient to reproduce the results reported in the original work~\cite{skoge:2006}. Finite size analysis conducted for packings in $d=6$ further indicates that no significant changes are observed for $N$ up to 3000.

Simulations are also employed for testing the theoretically predicted distribution functions of contact spheres (Sec.~\ref{sec:lowdcorr}). For this task, we randomly generate positions for $z$ contact balls on the surface of a $d$-dimensional central ball, and keep only the configurations that present no overlap. This approach guarantees that the resulting configurations do not depend on the dynamical sequence of sphere addition.

\section{The method}
\label{sec:method}

Before obtaining a high-dimensional form for the mean-field theory, we briefly review the volume function approach and generalize it to arbitrary $d$. Note that a more complete discussion of the case $d=3$ can be found in Ref.~\cite{song:2008,Song2010}.

\subsection{Calculation of the volume function}

\begin{figure}
\includegraphics[width=0.7\columnwidth]{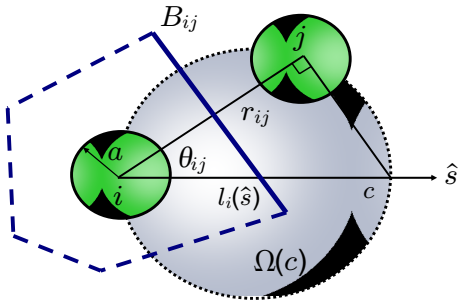}
\caption{
(Color online) Voronoi construction. The boundaries of the Voronoi cell of particle $i$ are illustrated by the
thicker lines. The Voronoi boundary $B_{ij}$ between the central particle $i$
and a Voronoi particle $j$ is a $(d$$-$$1)$-dimensional face bisecting the separation $r_{ij}$. The Voronoi
boundary along the direction $\hat{s}$ is given by $l_i(\hat{s})=r_{ij}/(2\cos \theta_{ij})$, where $\theta_{ij}$ is the
angle between $\hat{s}$ and $\hat{r}_{ij}$. The short-dashed (blue) sphere is the region $\Omega(c)$, where $c = 2 l_i(\hat{s})=r_{ij}/\cos \theta_{ij}$.
} \label{voronoi}
\end{figure}

Recall that a Voronoi cell is defined as a convex
polygon whose interior consists of the points that are closer to
a given particle than to any other. We refer to a
particle $j$ as the Voronoi particle of particle $i$ if $i$ and $j$ share a common Voronoi boundary $B_{ij}$, defined as the
$(d$$-$$1)$-dimensional surface that bisects the separation $r_{ij}$
between the two particles (Fig.~\ref{voronoi}). For
a given direction $\hat{s}$, the distance $l_i(\hat{s})$ from
particle $i$ to the boundary $B_{ij}$ is
\begin{equation}
l_i(\hat{s})=\min_{\hat{s} \cdot
\hat{r}_{ij'}}\frac{r_{ij'}}{2\hat{s} \cdot
\hat{r}_{ij'}}=\frac{r_{ij}}{2\hat{s} \cdot
\hat{r}_{ij}}=\frac{r_{ij}}{2 \cos \theta_{ij}}=\frac{c}{2},
\end{equation}
where $c \equiv
r_{ij}/{\cos \theta_{ij}}$. Operationally, a Voronoi particle $j$ is thus one that minimizes
$r_{ij'}/({2\hat{s} \cdot \hat{r}_{ij'}})$. Note that $c$ is the diameter of a spherical region
\begin{equation}
\Omega \left( c \right) \equiv \{ (r,\theta_1, \theta_2, \ldots
\theta_{d-1}) \ \ | \ \ r/\cos \theta_1 \leq c\},
\label{omega}
\end{equation}
where $(r,\theta_1, \theta_2, \ldots
\theta_{d-1})$ are $d$-dimensional spherical coordinates
and $\theta_1$ is the angle between $\hat{r}$ and $\hat{s}$  (see Fig.~\ref{voronoi}). If particle $j$ is truly a Voronoi particle, $\Omega (c)$ should be empty given that the central particle $i$ is at the origin.
The Voronoi volume of particle $i$ can thus
be expressed as the angular integral
\begin{equation}
W_i = \int d\hat{s} \, \frac{l_i(\hat{s})^d}{d}.
\end{equation}

It follows that the ensemble average $\langle \cdots \rangle$ of the total volume $V$ of a packing
\begin{equation}
\begin{split}
\langle V \rangle &= \left\langle \sum_{i=1}^{N} W_i \right\rangle =\left\langle \sum_{i=1}^{N} \int d\hat{s} \frac{l_i(\hat{s})^d}{d} \right\rangle
\\
&=NV_d \langle l_i(\hat{s})^d \rangle = N \langle W_i \rangle,
\end{split}
\label {eq:avev}
\end{equation}
where the last line results from assuming that the system is isotropic and homogenous,
such that the ensemble average of particle $i$ and direction $\hat{s}$ is equivalent to the overall ensemble average of the packing.
The reduced free volume per particle then simplifies to
\begin{equation}
w =\frac{ \langle W_i \rangle -V_g}{V_g}=\frac{V_{d}\langle l_i(\hat{s})^d \rangle-V_g}{V_g}=\langle c^d
\rangle - 1 \label{w} \ .
\end{equation}

The key quantity missing in the analysis of the Voronoi volumes is the probability distribution for $c$. We define
$f(c)dc$ as the probability that $l_i(\hat{s}) \in [c/2, (c+dc)/2]$ for $i$ at
the origin.
Note that because for hard spheres $c \in [1,\infty)$, by definition, $f(c)$ can only be non-zero
over this same interval.
We also define the inverse cumulative distribution function $P_>(c)$ that $\Omega(c)$ is empty of particle centers
for $i$ at the origin
\begin{equation}
P_>(c)\equiv 1-\int_1^c f(c') dc' \ ,
\end{equation}
and thus
\begin{equation}
f(c) = -\frac{dP_>(c)}{dc} \ .
\end{equation}
We then obtain that
\begin{equation}
\begin{split}
w = \frac1\phi -1  &=\int_1^\infty \left(c^d-1\right)f(c) dc \\
&=- \int_1^\infty \left(c^d-1\right)   \frac{dP_>(c)}{dc} dc   \\
 &= d\int_1^\infty c^{d-1}
P_>(c)dc,
\end{split}
\label{eq:sce}
\end{equation}
where the last line is obtained after integrating by parts and noting that the boundary terms vanish.
Tests of this identity on amorphous hard spheres packings
from simulations, where $P_>(c)$ is obtained directly from each packing (see Fig.~\ref{fig:factorization} below), confirm
the validity of the underlying isotropy assumption.

\subsection{Liquid state derivation of $P_>(c)$}
From Eq.~(\ref{eq:sce}), the problem of identifying the packing fraction at jamming is transformed into that of identifying the form of $P_>(c)$ from the structure of jammed configurations. The language of liquid state theory is particularly well-suited for this task because the jammed packings have structural features similar to that of high-density liquids.

Consider the $N$-particle probability density
$P_N(\mathbf{R}^N)$ of finding the particles $1,2,\ldots,N$ with configuration
$\mathbf{R}^N\equiv{\mathbf{R}_1,\mathbf{R}_2,\ldots,\mathbf{R}_N}$. Unit normalization is set by integrating the particle positions over space
\begin{equation}
\int P_N(\mathbf{R}^N)d\mathbf{R}^N = 1.
\end{equation}
The configurational average (or ensemble average) of a many-body observable
$F(\mathbf{R}^N)$ is then
\begin{equation}
\langle F(\mathbf{R}^N) \rangle \equiv \int F(\mathbf{R}^N)
P_N(\mathbf{R}^N) d\mathbf{R}^N.
\end{equation}
The associated reduced $n$-particle probability density (or $n$-point
correlation function)
\begin{equation}
\begin{split}
\rho_n(\mathbf{R}^n)& \equiv\!\!\!\!\!\! \sum_{i_1\neq i_2\neq \cdots
\neq i_n}^{\infty}\!\!\!\!\!\! \left\langle\delta(\mathbf{R}_1-\mathbf{R}_{i_1})
\delta(\mathbf{R}_2-\mathbf{R}_{i_2})
\cdots
\delta(\mathbf{R}_n-\mathbf{R}_{i_n})\right\rangle\\
&=\frac{N!}{(N-n)!} \int P_N(\mathbf{R}^n,
\mathbf{R}^{N-n}) d\mathbf{R}^{N-n}
\end{split}
\end{equation}

is itself normalized to
\begin{equation}
\int \rho_n(\mathbf{R}^n) d\mathbf{R}^n = \frac{N!}{(N-n)!}.
\end{equation}
For a system with translational invariance we can also define the $n$-particle correlation function
\begin{equation}
g_n(\mathbf{R}_{12},\mathbf{R}_{13}\ldots \mathbf{R}_{1n})\equiv \rho_n(\mathbf{R}^n)/\rho^n,
\end{equation}
with normalization
\begin{equation}
\rho^{n-1} \int g_n(\mathbf{R}_{12},\mathbf{R}_{13}\ldots
\mathbf{R}_{1n}) d\mathbf{R}_{12}\cdots \mathbf{R}_{1n} =
\frac{(N-1)!}{(N-n)!}.
\end{equation}
The $n$-particle correlation function reduces to the pair correlation function (or radial distribution function) for the case $n=2$.

Following the strategy of Refs.~\cite{Reiss:1959} and \cite{Torquato:1990} for expressing $P_>$ in terms of $n$-particle correlation functions, we define
\begin{equation}
m(\mathbf{R};\Omega)\equiv
\begin{cases}
1, \mathbf{R} \in \Omega\\
0, \text{otherwise},
\end{cases}
\end{equation}
as a characteristic function of space point $\mathbf{R}$ inside an arbitrary region $\Omega$. Here, we will specialize to the region $\Omega(c)$ defined by Eq.~(\ref{omega}). We also define the characteristic function
\begin{equation}
J(\mathbf{R}_1; \Omega)\equiv
\prod_{i=2}^{N}[1-m(\mathbf{R}_i-\mathbf{R}_1;\Omega)]
\end{equation}
of all $N-1$ particles outside region $\Omega$ centered at $\mathbf{R}_1$, with $\mathbf{R}_1$ the center of particle $1$. Considering the cumulative probability function
$P_>(\mathbf{R}_1;\Omega)$ as the probability that all $N-1$
particles are outside of $\Omega$
gives
\begin{widetext}
\begin{equation}
\begin{split}
P_>(\mathbf{R}_1;\Omega) &= \frac{N}{\rho_1(\mathbf{R}_1)} \int
J(\mathbf{R_1}; \Omega)P_N(\mathbf{R}^N)d \mathbf{R}^{N-1}\\
&=1-\frac{1}{\rho_1(\mathbf{R}_1)}\int
m(\mathbf{R}_2-\mathbf{R}_1;
\Omega)\rho_2(\mathbf{R}_1,\mathbf{R}_2)d \mathbf{R}_2\\
&+\frac{1}{2\rho_1(\mathbf{R}_1)}\int m(\mathbf{R}_2-\mathbf{R}_1;
\Omega)m(\mathbf{R}_3-\mathbf{R}_1;
\Omega)\rho_3(\mathbf{R}_1,\mathbf{R}_2,\mathbf{R}_3)d
\mathbf{R}_2\mathbf{R}_3-\cdots\\
&\equiv \sum_{k=0}^{N-1}(-1)^k F_k(\mathbf{R}_1;\Omega),
\end{split}
\label{Pg}
\end{equation}
where the last expression implicitly defines
\begin{equation}
F_k(\mathbf{R}_1;\Omega)\equiv
\begin{cases}
1 \ , \hskip1cm k=0 \ ,\\
\frac{1}{\rho_{1}(\mathbf{R}_1)k!}\int
\rho_{k+1}(\mathbf{R}^{k+1})\prod_{i=2}^{k+1}m(\mathbf{R}_{1i};\Omega)d\mathbf{R}_i \ , \hskip1cm k\geq 1 \ .
\end{cases}
\end{equation}
\end{widetext}
If the system has translational invariance, then
for $k\geq 1$
\begin{equation}
\begin{split}
F_k(\Omega)&=\frac{\rho^k}{k!}\int
g_{k+1}(\mathbf{R}_{12},\ldots,\mathbf{R}_{1(k+1)}
)\prod_{i=2}^{k+1}m(\mathbf{R}_{1i};\Omega)d\mathbf{R}_{1i}\\
&=\frac{\rho^k}{k!}\int_{\Omega}
g_{k+1}(\mathbf{R}_{12},\ldots,\mathbf{R}_{1(k+1)}
)d\mathbf{R}_{1i}\cdots d\mathbf{R}_{1(k+1)}.
\end{split}
\label{F}
\end{equation}
$F_2(\Omega)$, $F_3(\Omega)$, $\ldots$, are respectively the
probabilities of finding a pair, triplet, etc., within $\Omega$. Equations~(\ref{Pg})~and~(\ref{F}) show that the problem can be solved exactly only if a complete knowledge of the $n$-particle correlation function $g_n(\mathbf{R}_{12},\mathbf{R}_{13}\ldots \mathbf{R}_{1n})$ is available to all orders. But an accurate theoretical prediction of $g_n$ for jammed states is still lacking, even for the lowest order pair correlation function $g_2$. In the following, we use the generalized Kirkwood
superposition approximation~\cite{kirkwood:1935} and a theoretically conjectured form for the pair correlation function in high dimensions to simplify Eq.~(\ref{Pg}). These approximations result in a simple factorized form of $P_>(c)$ that only depends on the packing density $\rho$ and coordination number $z$.


For dimensions greater than one, the generalized Kirkwood
superposition approximation offers a way to
reexpress these higher-order correlations in terms of pair
correlations
\begin{equation}
g_n(\mathbf{R}_{12},\mathbf{R}_{13},\ldots, \mathbf{R}_{1n})\simeq
\prod_{1\leq i<j \leq n} g_2(\mathbf{R}_{ij}).
\end{equation}
In order to proceed any further with this analysis, we need to
approximate $g_2(\mathbf{r})$. Following Torquato and Stillinger's
suggestion~\cite{torquato:2006} and the results of replica theory~\cite{PZ10},
we postulate that the pair
correlation for a jammed configuration with $z$ average contacts
per particle is angularly independent and decomposable into
contact and bulk contributions as
\begin{equation}
g_2(r)\simeq\frac{z}{\rho S_{d-1}}\delta(r-1)+\Theta(r-1),
\label{eq:g2_torquato}
\end{equation}
where $\Theta(x)$ is the Heaviside step function.

The superposition approximation then becomes
\begin{widetext}
\begin{equation}
\begin{split}
g_n(\mathbf{R}_{12},\mathbf{R}_{13},\ldots, \mathbf{R}_{1n})&
\simeq \prod_{i=2}^{n}g_2(\mathbf{R}_{1i})\prod_{2\leq j<k \leq
n}g_2(\mathbf{R}_{jk})\\
& \approx \prod_{i=2}^{n}g_2(\mathbf{R}_{1i})\prod_{2\leq j<k \leq
n}[1+\frac{z}{\rho S_{d-1}}\delta(R_{jk}-1)-\Theta(1-R_{jk})]\\
& = \prod_{i=2}^{n}g_2(\mathbf{R}_{1i}) \left[1+\sum_{2\leq j<k
\leq n} \frac{z}{\rho S_{d-1}}\delta(R_{jk}-1)- \sum_{2\leq j<k
\leq n} \Theta(1-R_{jk})+\cdots \right].
\end{split}
\end{equation}
The relative importance of the various terms in the bracket comes
from the scaling of their integral over the volume $V$ of the spherical region
$\Omega$ with diameter $c$ and noting that $V_g/V\sim c^{-d}$
\begin{equation}
\begin{split}
\int_{\Omega}\left[1+\sum_{2\leq j<k \leq n} \frac{z}{\rho
S_{d-1}}\delta(R_{jk}-1)- \sum_{2\leq j<k \leq n}
\Theta(1-R_{jk})+\cdots \right]d\mathbf{R}_{12}\ldots
d\mathbf{R}_{1n} \\
=V^{n-1}+V^{n-2}\frac{(n-1)(n-2)}{2}\frac{z}{\rho}-V^{n-2}\frac{(n-1)(n-2)}{2}V_g+\ldots=V^{n-1}\left[1+\mathcal{O}\left(\frac{z}{\r V}\right)+\mathcal{O}\left(\frac{1}{c^d}\right)\right].
\end{split}
\end{equation}
\end{widetext}

\begin{figure}
\vbox{\includegraphics[width=.6\columnwidth]{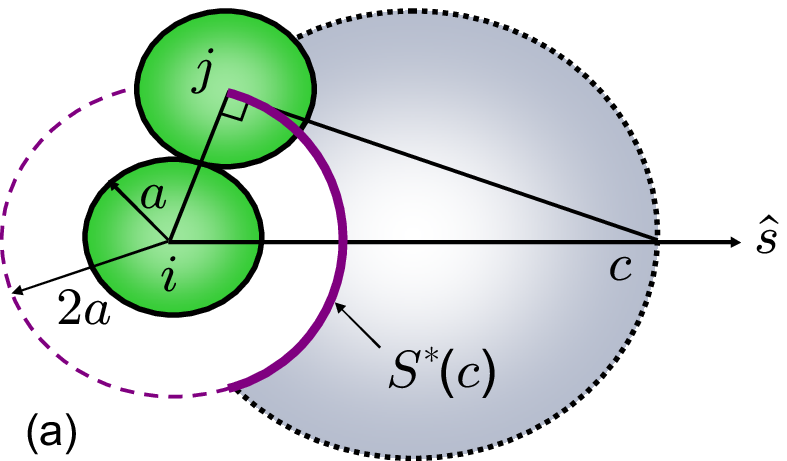}}
\vspace{0.5cm}
\vbox{\includegraphics[width=.6\columnwidth]{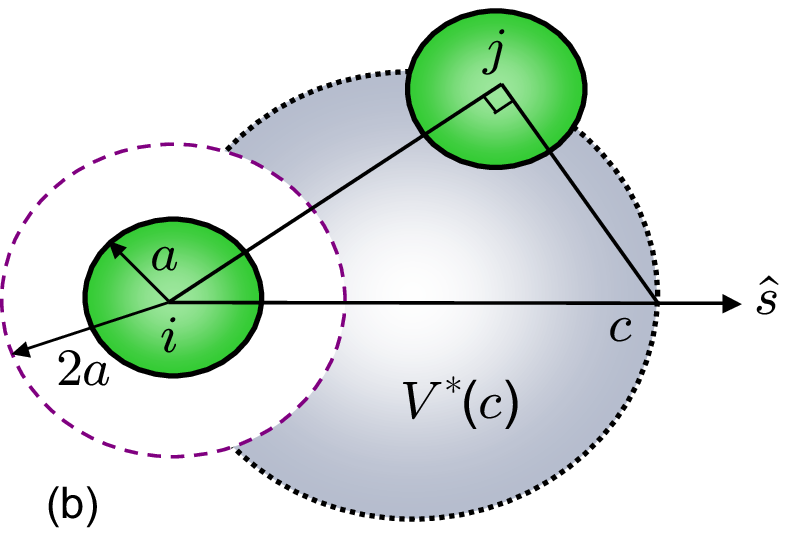}}
\caption{
(Color online) Schematic illustration of (a) $S^*(c)$ and (b) $V^*(c)$. The white sphere is the excluded zone of the central particle.
} \label{fig:sstar_vstar}
\end{figure}

To first-order we can then make the approximation
\begin{equation}
g_n(\mathbf{R}_{12},\mathbf{R}_{13},\ldots, \mathbf{R}_{1n})
\simeq \prod_{i=2}^{n}g_2(\mathbf{R}_{1i}),
\end{equation}
which amounts to saying that spheres $2...n$ are correlated
with the central sphere $1$ but not with each other. Though crude,
this treatment is reasonable in large $d$, because the sphere
surface is very large compared to the occupied surface and the
packing is increasingly inefficient. Plugging this result into
Eq.~(\ref{F}) gives
\begin{equation}
F_k(\Omega)=\frac{\rho^k}{k!}\left(
\int_{\Omega}g_2(\mathbf{R}_{12})d\mathbf{R}_{12} \right)^k
\end{equation}
and in the limit $N\rightarrow \infty$
\begin{align}
P_>(\Omega)&=\sum_{k=0}^{N-1}(-1)^k\frac{\rho^k}{k!}\left(
\int_{\Omega}g_2(\mathbf{r})d\mathbf{r} \right)^k\nonumber\\ &\approx
\exp\left[-\rho \int_{\Omega} g_2(\mathbf{r})d\mathbf{r}\right].
\label{pg2}
\end{align}
Approximating the  pair correlation with
Eq.~(\ref{eq:g2_torquato}) finally gives a factorized form whose
validity should improve with increasing dimension
\begin{align}
P_>(c) &= \exp \left[ - \rho V^*(c) -\frac{z S^*(c)}{S_{d-1}} \right]\nonumber\\
&=P_B(c)P_C(c), \label{eq:liqPgreater}
\end{align}
where
\begin{equation}
P_B(c)=\exp \left[-\rho V^*(c)\right]
\label{eq:liqPB}
\end{equation}
and
\begin{equation}
P_C(c)=\exp \left[-\frac{z S^*(c)}{S_{d-1}}\right]
\label{eq:liqPC}
\end{equation}
represent the contributions from bulk and contact particles respectively, and
\begin{align}
S^*(c)&=\int \delta(r-1)\Theta(c-r/\cos\theta)d\mathbf{r}\nonumber\\
&= S_{d-2} \int_0^{\arccos(1/c)}  d\theta (\sin \theta)^{d-2}
\end{align}
and
\begin{align}
V^*(c)&=\int \Theta(c-r/\cos\theta)d\mathbf{r}\nonumber\\
&=\frac{S_{d-2}}d \int_0^{\arccos(1/c)}  d\th (\sin \theta)^{d-2}
[ (c \cos\theta)^d -1 ].
\end{align}
are empty of particle centers (see Fig.~\ref{fig:sstar_vstar}). Fig.~\ref{fig:sstar_vstar} (b) also shows that $V^*(c)$ is smaller than the volume of $\Omega(c)$, because of the volume exclusion by the central particle.

\section{Large $d$ analytical solution}
\label{sect:analphi}

Substituting the factorized form Eq.~(\ref{eq:liqPgreater}) for $P_>(c)$ into Eq.~(\ref{eq:sce}),
one obtains a self-consistency relation for $w$ that is amenable to further analytical and numerical treatment in the high-dimensional limit
\begin{align}
w & = d \int_1^\io dc \, c^{d-1}  \exp \left[ - \frac{2^d V^*(c)}{ (w+1) V_d } -\frac{z S^*(c)}{S_{d-1}} \right].
\label{weq}
\end{align}

\subsection{Change of variable}
\label{subsect:vari}

For notational convenience, we define
\begin{align}
s^*(1/c) \equiv& \int_0^{\arccos(1/c)}  d\th (\sin \th)^{d-2} \nonumber\\ =& \int_{1/c}^1 d\xi (1-\xi^2)^{(d-3)/2}
\label{eq:sstar}
\end{align}
and
\beq\begin{split}
v^*(1/c) \equiv & \int_0^{\arccos(1/c)}  d\th (\sin \th)^{d-2}  [ (c \cos\th)^d -1 ]  \\
 =& \frac{c^d}{2^d} \left[ s^*(-1) - s^*(1-2/c^2) \right] +\\
 & \frac{c}{2d-2} \left( 1 - \frac{1}{c^2} \right)^{\frac{d-1}{2}} - s^*(1/c),
\end{split}\eeq
and note that $s^*(-x) = s^*(-1) - s^*(x)$. We reexpress
\beq\begin{split}
& \frac{V^*(c)}{V_d}
= \frac{S_{d-2}}{S_{d-1}} v^*(1/c)
= \frac{ v^*(1/c) }{s^*(-1)} \\
&\frac{S^*(c)}{S_{d-1}}
= \frac{S_{d-2}}{S_{d-1}} s^*(1/c) = \frac{s^*(1/c)}{s^*(-1)}.
\end{split}\eeq
For large $d$ and $x>0$ we can evaluate $s^*(x)$ by saddle-point approximation. Developing the exponential around $\xi = x$ and using the fact that the integrand decays rapidly in large $d$ allows to extend the upper boundary of integration to $\xi = \infty$ and leaves corrections of order $\mathcal{O}(1/d)$, i.e.,
\beq\label{eq:sstarsaddle}
\begin{split}
s^*(x)  &= \int_{x}^1 d\xi (1-\xi^2)^{(d-3)/2}\\
&= \int_x^1 d\xi \, e^{\frac{d-3}2 \log(1-\xi^2)} \\
&\approx
 \int_x^\io d\xi \, e^{\frac{d-3}2 \{ \log(1-x^2) - \frac{2x}{1-x^2} (\xi - x) - \frac{1+x^2}{(1-x^2)^2} (\xi - x)^2 \} } \\
 &=\frac{e^{\frac{(d-3) x^2}{2 (1+x^2)}} (1-x^2)^{d/2}}{\sqrt{\frac{2}{\pi}(d-3) (1-x^4)}} \left[1 - \textrm{Erf}\left( \sqrt{  \frac{(d-3) x^2}{2 (1+x^2)}}  \right) \right]
\\&\simeq \frac{1}{d x}(1-x^2)^{(d-1)/2},
\end{split}\eeq
where $\textrm{Erf(x)}$ is the error function. Note that the last expression is only reasonable for $x$ away from zero.

By using the relations above and noting that $w\sim w+1$ in the high-dimensional limit, Eq.~(\ref{weq}) becomes
\beq
w = d \int_1^\io dc \, c^{d-1} \exp \left[   - \frac{c^d}{w} + H_d(c;w,z)   \right]
\eeq
with
\begin{widetext}
\beq
\begin{split}
H_d(c;w,z) & =
\frac{1}{s^*(-1)} \left\{ \frac{c^d}{w} s^*(1-2/c^2) -
\frac{2^d c (1-1/c^2)^{(d-1)/2}}{w (2d -2)} + \left(\frac{2^d}w - z \right) s^*(1/c) \right\} \\
& \simeq \frac{2^d}{w s^*(-1)} \left(1 - \frac{1}{c^2}\right)^{(d-1)/2}
\left\{ \frac{c}{2d (1-2/c^2)} - \frac{c}{2d-2} + \frac{c}d \left( 1 - \frac{z w }{2^d} \right) \right\},
\end{split}\eeq
\end{widetext}
where the last result is obtained by using the simplified expression for $s^*(x)$ in Eq.~(\ref{eq:sstarsaddle}). A change of variable $y = c^d/w$ further reduces the expression to
\beq
1 = \int_{1/w}^\io dy \, e^{-y + H_d[ (w y)^{1/d} ; w,z ]}.
\eeq
Because we expect $w$ to exponentially diverge for large $d$, the lower integration limit should rapidly go to zero. For any finite $y$, $y^{1/d} \to 1$ in high dimension. The equation above then becomes
\beq
1 \sim  \int_{0}^\io dy \, e^{-y} e^{H_d( w^{1/d} ; w,z )} = e^{H_d(w^{1/d};w,z)} \ .
\label{H}
\eeq
Assuming $H_d$ is well behaved at the boundaries, the problem reduces to finding a solution for the condition
\begin{equation}
H_d(w^{1/d};w,z) = 0.
\label{eq:condition}
\end{equation}

\begin{figure}[t]
\includegraphics[width=\columnwidth]{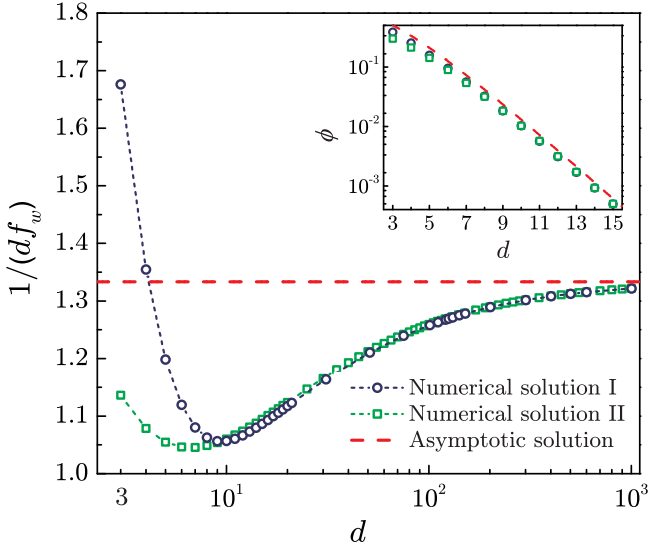}
\caption{(Color Online) The values of $1/(d \, f_w)$ obtained from (i) the exact
numerical solution of Eq.~(\ref{weq}) (numerical solution I), (ii) the correction
to the asymptotic analysis in large $d$ obtained by numerically solving 
Eq.~(\ref{eq:condition_iso}) (numerical solution II), and (iii) the leading term Eq.~(\ref{f_w}) in
the asymptotic analysis (asymptotic solution). The insert shows the dimensional scaling of the corresponding volume fraction $\phi$ from Eq.~(\ref{eq:phiScale}). The numerical values for $d=3-6$ are also listed in Table~\ref{tab:phij}.}
\label{fig1}
\end{figure}

\subsection{Asymptotic solution}
\label{subsect:solution}

For analytical convenience we pose that $z$ and the solution $w(z)$ have a form
\beq\begin{split}
&w = f_w w_0^d \\
&z = f_z z_0^d,
\end{split}\eeq
where $f_w$ and $f_z$ are polynomial functions of $d$. Equation~(\ref{eq:condition}) then becomes
\beq
\frac{w_0}{2d (1-2/w_0^2)} - \frac{w_0}{2d-2} + \frac{w_0}d \left[ 1 - f_z f_w \left(\frac{z_0 w_0}2 \right)^d \right] =0.
\eeq
The necessity to avoid an exponential divergence of the last term imposes
\begin{equation}
w_0 = 2/z_0,
\label{eq:w0z0}
\end{equation}
and allows to express $f_w$ as a function of $f_z$. In the isostatic case $z = 2d$, i.e., when $z_0=1$ and $f_z(d) = 2d$, simple algebraic manipulations give $f_w =\frac{3}{4d}$, or equivalently,
\begin{equation}
w =  \frac{3}{4d}2^d,
\label{f_w}
\end{equation}
and
\begin{equation}
\phi \sim  \frac{4d}{3}2^{-d}.
\label{eq:phiScale}
\end{equation}

The scaling form of $w$ provides a way to critically examine the simplifications done to  $s^*(x)$. Using the rigorous bounds on packing volume fraction for large $d$, $2^{-d} \leq \phi \leq 2^{-0.5990 d}$~\cite[Chap 1.
Sec. 1.5]{conway:1999},
and $w \sim 1/\phi$, we get $2 \geq w_0 \geq 2^{0.5990} \approx 1.5146$. Because $w^{1/d} \to w_0$, we use the bound on $w_0$ to directly check that
$1/w_0 \in [0.5,0.66]$ and $1-2/w_0^2 \in [0.128,0.5]$. We conclude that the use of Eq.~(\ref{eq:sstarsaddle}) to approximate $s^*(x)$ in $H_d$ is justified, because its arguments are always positive and bounded away from zero. The validity of the simplified expression for $H_d$  in Eq.~(\ref{H}) is also verified.

\subsection{Finite $d$ corrections}
\label{subsect:finiteD}

Assuming that Eq.~(\ref{eq:condition}) has only corrections that are subdominant, the above analysis also provides the leading finite $d$ corrections to $w$. These corrections can be obtained by numerically solving the condition $H_d(c;w,z)$ with the exact expression Eq.~(\ref{eq:sstarsaddle}) for $s^*$ in the case $z=2d$ and $w = f_w 2^d$, then computing $f_w$ from the condition
\beq
H_d(2 f_w^{1/d};f_w 2^d,2d) = 0.
\label{eq:condition_iso}
\eeq
The results are compared with the self-consistent solution of Eq.~(\ref{weq}) obtained by direct numerical evaluation of the expressions
with the exact form for $s^*$ in Fig.~\ref{fig1}. The quality of the match at high $d$ provides a further verification of the simplifying approximations made in deriving Eq.~(\ref{eq:condition}).

Both treatments agree rather well down to $d\approx 10$. The change in behavior of the packing fraction then observed corresponds to where the decay of $P_>(c)$
becomes slower than the growth of $c^{d-1}$ as the dimension increases.
This inversion results in $c\sim 2$ maximizing the integrand in high dimensions instead of $c\sim 1$, as in low dimensions. Eq.~(\ref{eq:condition})
is derived under the assumption that the integral is dominated
by $c\sim 2$, which may explain why the two curves deviate from each other in that dimensional regime.
Physically, this change corresponds to the jammed packings becoming relatively less efficient with dimension. In high dimensions, nearby spheres do not provide sufficient cover of the sphere surface, which results in the inclusion of farther neighbors within the Voronoi cell. In low dimensions, the screening is efficient, so the integral is dominated by low $c$ values. Because an open structure is less prone to particle-particle correlations, this interpretation is at least consistent with the assumptions made in developing the theory.

\section{Low $d$ analysis}
\label{sec:lowd}

The analysis in the previous section is based on the approximate high-dimensional expression for $P_>(c)$ from Eq.~(\ref{eq:liqPgreater}) .
Errors resulting from the superposition approximation and the
postulated form for $g_2(r)$ should be taken into account, in order to
improve $\phi$ results in finite dimensions. Though a systematic expression for
these corrections is not trivial to obtain, it is
nonetheless possible to guess some of their forms by inspection. And by comparing their predictions with the properties of packings obtained from simulation, we can assess their success. Our low-dimensional analysis expands and generalizes the approach of Ref.~\cite{song:2008} for the case $d=3$~\cite{song:2008,Song2010}.

\subsection{Low $d$ corrections of $P_>(c)$}
\label{sec:lowdcorr}
\begin{figure}
\centerline{\includegraphics[width=\columnwidth]{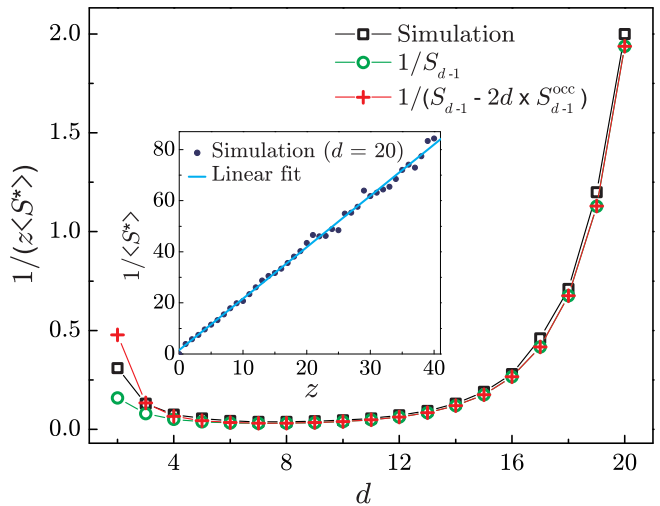}}
\caption{
(Color Online) Comparison of the approximations for $\langle S^*\rangle$
from Eq.~(\ref{eq:rhoShigh})  and
Eq.~(\ref{eq:rhoSlow}) with the simulated
results.
The results show that $1/ \langle S^*\rangle $ is proportional to $z$ when $z$ is not too large ($z\leq2d$).
(inset) The simulation results in $d=20$ illustrate the linear behavior of $1/\langle S^* \rangle$ with $z$.}
\label{fig:sstar}
\end{figure}

\begin{figure}
\includegraphics[width=\columnwidth]{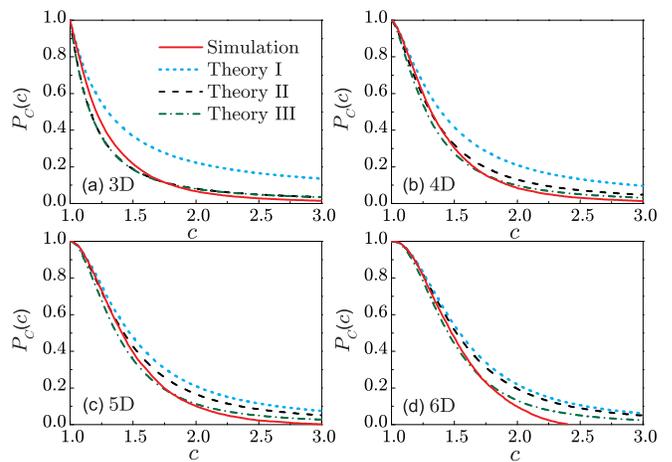}
\caption{(Color Online) Comparison of the simulation and theoretical
forms of $P_C(c)$ from Eq.~(\ref{eq:pc}) using different $\langle S^*\rangle$:
(i) Eq.~(\ref{eq:rhoShigh}) for a low-density (high-dimensional) limit (theory I),
(ii) Eq.~(\ref{eq:rhoSlow}) for a van der Waals-like correction (theory II), and (iii) direct surface simulations of $\langle S^*\rangle$ (theory III).}  \label{fig:PCc}
\end{figure}

\begin{figure}
\includegraphics[width=\columnwidth]{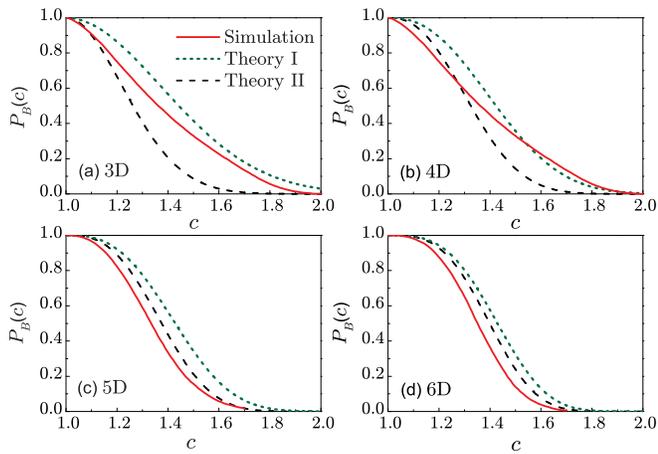}
\caption{(Color Online) Comparison of the simulation and theoretical
forms of $P_B(c)$ from Eq.~(\ref{eq:pb}) using the packing
fraction from simulation and different $\langle V^* \rangle$: (i) Eq.~(\ref{eq:rhohighD}) for the low density limit (theory I) and (ii)
Eq.~(\ref{eq:rholowD}) for a van der Waals-like correction (theory II). } \label{fig:PBc}
\end{figure}

\begin{figure}
\centerline{\includegraphics[width=\columnwidth]{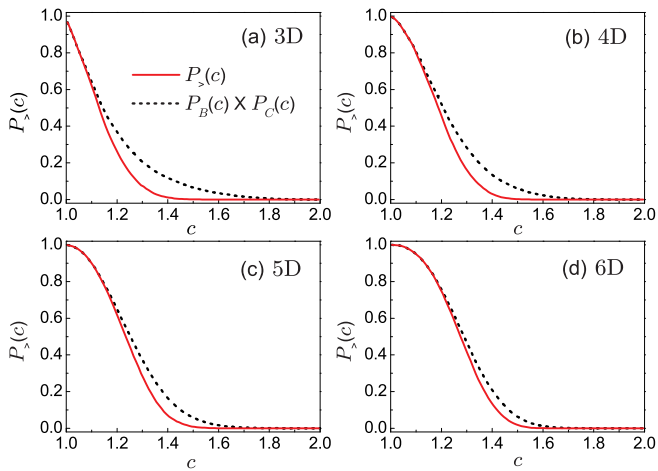}}
\caption{(Color Online) Evaluation of the factorization approximation by
comparing simulation results for $P_>$ in jammed systems with the
product of the bulk and contact contributions also from
simulations.} \label{fig:factorization}
\end{figure}

We first assume that $P_>(c)$ can be factorized in bulk $P_B(c)$ and contact $P_C(c)$ contributions as
in Eq.~(\ref{eq:liqPgreater}), and
examine the correlations between the different types of particles separately. The
contact term has the general form
\begin{equation}
P_C(c)=e^{-S^*(c)/\langle S^*\rangle}, \label{eq:pc}
\end{equation}
because
\begin{equation}
\begin{split}
\langle S^*\rangle &=\int_1^\infty S^*(c)f(c)dc=\int_0^1 S^* \, dP_C\\
 &\approx \int_0^\infty P_C \, dS^*,
\end{split}
\end{equation}
where the last step is obtained after integrating by parts and assuming that the upper limit of the integration (the maximum of $S^*$) goes to infinity. The average of the available solid angle $\langle S^*\rangle$ can also be calculated directly from simulations of surface sphere configurations. In the limit of low occupancy, where the particle volume is negligible
and correlations are absent, the probability that $z$ contact particles lie outside $S^*(c)$ is
\begin{equation}
P_C(c)\approx\left[1-\frac{S^*(c)}{S_{d-1}}\right]^z \approx \exp\left[-\frac{zS^*(c)}{S_{d-1}}\right]
\end{equation}
or
\begin{equation}
\langle S^*\rangle\approx\frac{S_{d-1}}{z}. \label{eq:rhoShigh}
\end{equation}
This result is the same as Eq.~(\ref{eq:liqPC}) from the previous
analysis, because an equivalent approximation to the low-density approximation was taken in the high-dimensional treatment~\cite{torquato:2006}.
Surface correlations are, however, particularly
significant in low $d$ even for low occupancy, and therefore the $S_{d-1}$ volume (sphere
surface) should be corrected for the non-negligible volume occupied by the particles. We define the surface occupied by a sphere at contact
\begin{equation}
S^{\mathrm{occ}}_{d-1}\equiv S_{d-2}\int_0^{\pi/6}(\sin{\theta})^{d-2}d\theta,
\end{equation}
to include a van der
Waals-like correction for the occupied volume
\begin{equation}
\langle S^*\rangle\approx\frac{S_{d-1}-z
S^{\mathrm{occ}}_{d-1}}{z}. \label{eq:rhoSlow}
\end{equation}
This expression reduces to the decorrelated form in high dimensions,
because $S^{\mathrm{occ}}_{d-1}/S_{d-1}$ vanishes faster than
$z\sim d$ grows. Figure~\ref{fig:sstar} shows that the latter is slightly better at low $d$ and that both
approximations rapidly converge to the $\langle S^*\rangle$
numerically obtained from random distributions of surface sphere
configurations.

A similar treatment can be applied to the bulk term, which has the form
\begin{equation}
P_B(c) = e^{- V^*(c)/\langle V^*\rangle} \label{eq:pb}
\end{equation}
with $\langle V^*\rangle$ the average volume available to
particles beyond the contact rim. Without taking into account the
volume excluded by the central sphere, and assuming that there
are no angular correlation, the large $d$ limit is recovered
\begin{equation}
\langle V^*\rangle=V/N=1/\rho. \label{eq:rhohighD}
\end{equation}
A correction \emph{\`a la} van der Waals for the correlations that arise from the excluded
volume by the central particle gives a form similar to that of
the surface term
\begin{equation}
\langle V^*\rangle=\frac{V-N V_g}{N}= V_g w.
\label{eq:rholowD}
\end{equation}
The expression reduces to the uncorrelated form in high
dimensions, because the volume of a unit-diameter ball grows much slower than the volume per particle at jamming.

The contact and bulk corrections above are equivalent to those
obtained in the original derivation of the
theory in $d=3$~\cite{song:2008}. Comparing
the different expressions with the simulation results should inform us on the reasonableness of the approximations made in their derivation. As can be seen in
Fig.~\ref{fig:PCc}, even a crude treatment of volume exclusion
improves the quality of the scaling form of the contact
contribution~\cite{footnote:2}. Using the simulated value of $\langle
S^*\rangle$ does even better. Moreover, the agreement with the scaling form
steadily improves from three to six dimensions.
Figure~\ref{fig:PBc} shows that the bulk contribution is also
better captured when correlations are included. We note however that though the agreement of
the bulk scaling form improves with dimension, convergence is
slower than for the contact contribution. Higher-dimensional
comparisons would be useful to confirm the trend, but are
unfortunately beyond current computational reach.

Even assuming that we could obtain separate perfect bulk and contact scaling forms, the factorization of the two contributions itself
remains an approximation. The corrections to this approximation are not easily directly tractable, but the rapidity at
which it vanishes can be indirectly evaluated by
analyzing the structure of jammed configurations.
Figure~\ref{fig:factorization} indicates that
corrections to the factorization vanish fairly rapidly with dimension, which supports the validity of the high-dimensional treatment.

\subsection{Low $d$ jammed packing fraction}
\begin{table}
\begin{ruledtabular}
\begin{tabular}{|c|c|c|c|c|}
  $d$ &
  $\phi_{\mathrm{sim}}$ & $\phi_{\mathrm{fact}}$ & $\phi_{\mathrm{lowD}}$ &  $\phi_{\mathrm{highD}}$ \\\hline
  3 & 0.64(1) & 0.54 & 0.64 & 0.39 \\
  4 & 0.46(1) & 0.39 & 0.38 & 0.25 \\
  5 & 0.31(1) & 0.27 & 0.22 & 0.16 \\
  6 & 0.20(1) & 0.18 & 0.13 & 0.10 \\
\end{tabular}
\end{ruledtabular}
\caption{Jamming packing fraction of frictionless spheres at
$z=2d$. Compression results from various simulation approaches $\phi_{\mathrm{sim}}$
(see text) are compared with the integration of the surface and
bulk decomposition of $P_>(c)$ ($\phi_{\mathrm{fact}}$), the low-dimensional approximation  ($\phi_{\mathrm{lowD}}$), and the high-dimensional
approximation ($\phi_{\mathrm{highD}}$), as introduced in the text.
} \label{tab:phij}
\end{table}

Comparing the numerical jamming packing fraction obtained under
different approximation schemes with the direct compression of the
system for various dimensions provides a final evaluation of
the theoretical treatment (see Table~\ref{tab:phij}). From a simulation perspective, it is reassuring to note that the packing
fraction obtained from compression simulations agree not only with the
previously reported 4-6d values~\cite{skoge:2006}, but also with the free volume
estimates from the metastable liquid state~\cite{vanmeel:2009},
the zero-shear limit in 4d~\cite{otsuki:2009}, and the canonical
3d results~\cite{bernal:1960,SK69,berryman:1983}.

From the theoretical point of view, three different levels of
approximation for $P_>$ are selected to solve Eq.~(\ref{eq:sce}).
\begin{itemize}
\item $\phi_{\mathrm{highD}}$: the high-dimensional approximation is the numerical solution to Eq.~(\ref{weq}) plotted in Fig.~\ref {fig1} (numerical solution I). This approximation is only valid in the high-dimensional or low-density limit. Therefore, $\phi_{\mathrm{highD}}$ is expected to have significant deviations from the simulation values in low $d$.

\item $\phi_{\mathrm{lowD}}$: the low-dimensional corrections Eqs.~(\ref{eq:pc}),~(\ref{eq:pb}) and (\ref{eq:rholowD}) to $P_>(c)$ give
\begin{align}
  P_>(c) \approx \exp \left[ - \frac{2^d V^*(c)}{ w V_d } -\frac{S^*(c)}{\langle S^* \rangle} \right],
\label{eq:PgLowD}
\end{align}
where $\langle S^*\rangle$ is obtained from analyzing simulations of surface sphere configurations~\cite{footnote:1}. This is exactly the same approach as what was used in Ref.~\cite{song:2008} for the 3D case.

\item $\phi_{\mathrm{fact}}$: the factorized approximation of $P_>(c) = P_B(c)P_C(c)$  is obtained by computing $P_B(c)$ and $P_C(c)$ directly from simulations of jammed configurations.
\end{itemize}

The first approximation is the crudest, while the last one only assumes
$P_>$ to be factorizable. The latter should therefore be the most accurate and become more so with increasing dimension. The factorized version $\phi_{\mathrm{fact}}$ indeed steadily approaches the simulation results $\phi_{\mathrm{sim}}$ with increasing dimension. The
agreement of $\phi$ with $\phi_{\mathrm{sim}}$ with
theoretical sophistication from $\phi_{\mathrm{highD}}$ to $\phi_{\mathrm{fact}}$ also generally improves for a given dimension. Note that the 3D result for
frictionless spheres of Song \emph{et al.}~\cite{song:2008} ($\phi_{\mathrm{lowD}}$ in Table~\ref{tab:phij}), which agrees very well with the simulation value, defies however this last trend.

Deviations between theory and simulation confirm that correlations are not
negligible in low dimensions. The assumption of the
factorizability of $P_>$ as well as the theoretical forms of $P_B$ and $P_C$ should thus be refined by including the
contributions from these correlations, in order to obtain a systematic quantitative agreement in low dimensions.

\section{Discussion and relation with other approaches}
\label{sec:comp}

In this section, we discuss the relation of the results obtained above with other approaches to the problem of sphere packing
in large dimension. The present theory predicts that
amorphous isostatic, i.e., random close packed, configurations have a \emph{unique} packing fraction $\phi \sim (4/3) \, d \, 2^{-d}$. This scaling
lies within the known rigorous upper and lower bounds for packings. But because these constraints are not so difficult to satisfy, it is more instructive to contrast the scaling form with
that of approaches developed specifically to deal with amorphous packings.

A first set of results for amorphous packings has been obtained by analyzing the behavior of a class of simple algorithms,
such as Ghost Random Sequential Addition (GRSA). GRSA is actually able to \emph{construct} packings up to density
$\phi=2^{-d}$, which provides a nice way of generating amorphous configurations up to the Minkowski
lower density bound~\cite{TS06}.
Random Sequential Addition (RSA),
where one attempts to add a sphere randomly and accepts the move only if there are no overlaps, produces more efficient packings. But RSA is already too complex to be analyzed analytically, so one has to resort to numerical
investigations~\cite{TTVV00,TUS06}. The results of RSA are consistent with $\phi = d \, 2^{-d}$, which is closer to but still lower than our scaling form.
It is well known that RSA algorithms are not very efficient in low dimensions.
A much more efficient approach is the Lubachevsky-Stillinger (LS) algorithm, which is able to construct isostatic packings with a density close to random close packing in low dimensions (see Sect.~\ref{sec:details})~\cite{skoge:2006}.
Unfortunately, investigating the LS algorithm in the large $d$ limit is not possible. The extrapolation of the scaling form to low
dimensions $\phi \sim 2.56 \, d \, 2^{-d}$~\cite{skoge:2006}
is probably not reliable due to the change in
the nature of packings around $d\approx 10$, as discussed above.

In the absence of a direct way to analyze the packing algorithms, we resort to an alternate approach that provides an estimate
of the random close packing density in large $d$. The mean field approach to the glass transition known as
Random First Order Transition (RFOT) theory~\cite{KW87}
assumes that amorphous packings correspond to the infinite pressure
limit of long-lived metastable hard sphere glasses~\cite{PZ10}. The problem of random close packing
is then reduced to the study of a simpler equilibrium problem, that of computing the equation of state of the glass.
This approach predicts that the system remains a liquid~\cite{FP99,PS00} up to a certain
packing fraction $\phi_{\mathrm{dyn}}$, where a dynamical transition to a finite pressure glass occurs~\cite{KW87}. Inside the glass phase,
a huge number of glassy states coexist. Applying an infinite pressure on these glassy states results in isostatic jammed configurations with a \emph{range} of packing fractions, as has been numerically verified in low dimensions~\cite{skoge:2006,CBS09}. In general, the range extends from a threshold (th) packing fraction to the glass close packed (GCP) packing fraction, i.e.,
$\phi \in [\phi_{\mathrm{th}},\phi_{\mathrm{GCP}}]$~\cite{PZ10}.
One of the key qualitative results of the RFOT treatment therefore differs from the present theory's uniqueness prediction.

Comparing the detailed high-dimensional behavior of the various approaches sheds some light on this disagreement. Several different implementations of the general RFOT approach provide scaling forms for its constitutive densities.
\begin{itemize}
\item Density Functional Theory (DFT) predicts $\phi_{\mathrm{dyn}} = \sqrt{2 \pi e} \, d \, 2^{-d} \sim 4.13 d \, 2^{-d}$~\cite{KW87}.
\item Mode-Coupling Theory (MCT) predicts $\phi_{\mathrm{dyn}} \sim d^2 \, 2^{-d}$ in its full version \cite{IM10,SS10} and
$\phi_{\mathrm{dyn}} = 2 \sqrt{2 \pi e} \, d \, 2^{-d} \sim 8.26 d \, 2^{-d}$ when using a Gaussian approximation for the non-ergodic parameter~\cite{KW87,IM10}.
It was shown in Ref.~\cite{IM10}, however, that MCT leads to inconsistencies above $\phi \sim d \, 2^{-d}$.
The scaling $\phi_{\mathrm{dyn}} \sim d^2 \, 2^{-d}$ predicted by the full MCT thus seems unreliable.
\item Replica Theory (RT) predicts $\phi_{\mathrm{dyn}}~\sim~4.8 \, d \, 2^{-d}$ for the dynamical transition, and $\phi_{\mathrm{th}}~\sim~6.26 \, d \, 2^{-d}$ and $\phi_{\mathrm{GCP}}~\sim~d \, \ln(d) \, 2^{-d}$
as boundaries for jamming.
\end{itemize}
Interestingly, DFT, RT, the corrected MCT, and the present theory all predict $\phi \sim d \, 2^{-d}$ for the glassy/jamming density.
It remains a problem of RFOT to reconcile the $\phi_{\mathrm{dyn}}$ predictions coming from the different approaches (DFT and MCT cannot make any prediction
for the jamming densities).
But, for now, all RFOT theories give $\phi_{\mathrm{dyn}}$ bigger than
$(4/3) \, d \, 2^{-d}$, which suggest that the system should still be a liquid at the density where the current theory predicts it to be jammed.
If one believes the RFOT results, this discrepancy suggests that the present theory might still be missing some correlations that, although irrelevant to determine the overall
scaling of the density, are important for the precise determination of the prefactor.
The replica theory prediction that jammed packings exist in an interval $\phi \in [6.26 \, d \, 2^{-d}, d \, \ln(d) \, 2^{-d}]$, whose width grows with dimension might be also encoded in some missing correlations.
Understanding the nature of these correlations, or disproving the replica results, could be considerable advances in the
microscopic comprehension of amorphous high- and low-dimensional jammed packings.

Finally, an interesting byproduct of the current analysis is a relation between the volume
fraction $\phi$ and the average coordination number $z$ in high dimensions from Eq.~(\ref{eq:w0z0})
\begin{equation}
\phi \sim \frac{z}{2^d}.
\end{equation}
This scaling is consistent with the results obtained in Refs.~\cite{torquato:2006,scardicchio:2008}, which use a rather different approach.
The relation thus seems well verified (possibly with sub-exponential corrections) for amorphous packings and a range of known
lattice (see \cite{conway:1999,SlWeb} for the volume fraction and the kissing number of a list of all known densest packings up to $d=128$).

\section{Conclusion}
\label{sec:conclusion}

In this paper, we have obtained an asymptotic high-dimensional density scaling of random close packings using a statistical theory. The scaling form is consistent with that obtained from other theories. Comparisons between the numerical simulations and the structural approximations support the theory's validity in high dimensions.

The theory provides a general method for relating the local surface
constraint in a jammed packing to its global properties. A simple relation between the volume fraction and the average coordination number is obtained by constraining the contact value of the pair distribution function. We note, however, that the current approach is a mean-field
theory that neglects unconstrained spatial correlations and
coordination number fluctuations.
The latter
might allow for better compactified packing
structures and the former may
exist in an amorphous packing of monodisperse spheres, even though
the system appears mescoscopically homogeneous.
The low-dimensional results suggest indeed that the role of spatial
correlations can be particularly significant.
Higher level coarse-graining, such as explicitly treating the second-layer
neighbors, the contributions from coordination number
fluctuations and
spatial correlations might thus improve the theory's predictions.

Though the current theory neglects three- and higher-body correlations as well as noncontact two-body correlations, the consistency of our scaling form with that of the other theoretical approaches discussed in Sec.~\ref{sec:comp} nonetheless suggests the presence of a certain universality in the packings of amorphous spheres in this limit, that is, the impact of spatial correlations on jammed packings may be greatly suppressed in high dimensions. If the high-dimensional asymptotic behavior is indeed related to the asymptotic behavior in the low-density limit for any finite $d$, as was proposed in Ref.~\cite{torquato:2006}, then dimensional studies could provide further constructive information on the nature of the jammed state.
To that end, the intrinsic inclusion of higher-order correlations in MCT, DFT, and RT appears advantageous, but the theories' built-in complexity also partly obscures the physical origin of these correlations.
The dimensional perspective provided by the current framework might thus be a promising geometrical starting point for identifying the nature and quantifying the corrections present in packings of any dimension. Our understanding of the experimentally accessible two- and three-dimensional packings would certainly benefit from such an advance.

Finally, though the scaling form we obtain gives amorphous packings that are denser than the Minkowski lower bound, it is still too far from the rigorous crystalline upper bound for bringing any resolution to the lattice vs. amorphous packing question.

\begin{acknowledgments}
We acknowledge useful discussions with Professor Hern\'an Makse, Professor Kunimasa Miyazaki and Professor Rolf Schilling. PC acknowledges Duke startup funding. FZ acknowledges hospitality at the Princeton Center for Theoretical Physics (PCTS) during part of this work.
\end{acknowledgments}


\begin{thebibliography}{65}
\bibitem[{foo({\natexlab{a}})}]{footnote:2}
\bibinfo{note}{Note that the agreement with the theoretical tail of $P_c(c)$
  improves with increasing simulated system size in $d=6$ (not shown), but does
  not significantly affect the agreement with $P_>(c)$.}
\end{thebibliography}
\end{document}